\documentclass{article}

\usepackage{epsfig}
\usepackage{graphics}
\usepackage{graphicx}
\usepackage[centertags]{amsmath}
\usepackage{amsfonts}
\usepackage{amsthm}
\usepackage{amssymb}
\usepackage{url}
\usepackage{subfigure}

\begin{document}

\title{On logical gates in precipitating medium:\\ cellular automaton model}

\author{Genaro Ju\'arez Mart\'{\i}nez$^1$ \\ Andrew Adamatzky$^1$ \\ Ben De Lacy Costello$^2$}

\date{}

\maketitle

\begin{centering}
$^1$ Faculty of Computing, Engineering and Mathematical Sciences, University of the West of England, Bristol, United Kingdom. \\
$^2$ Faculty of Applied Sciences, University of the West of England, Bristol, United Kingdom \\
\end{centering}

\begin{abstract}
\noindent
We study a two-dimensional semi-totalistic binary cell-state cellular automaton, which imitates a reversible precipitation in an abstract chemical medium. The systems exhibits a non-trivial growth and nucleation. We demonstrate how basic computational operation can be realized in the system when the propagation of the growing patterns is self-restricted by stationary localizations. We show that precipitating patterns of different morphology compete between each other and thus implement serial and non-serial logical gates. \\ %and majority gates.\\

\noindent
\textit{Keywords:} cellular automata, chemical reaction, growth, nucleation, waves, collisions and universality
\end{abstract}

\section{Introduction}

Constraining the media geometrically is a common technique used when designing computational schemes (particularly implementing logical circuits) in spatially extended non-linear media. For example `Strips' or `channels' are constructed within the medium (e.g. excitable medium) and connected together, typically using arrangements such as T-junctions. Fronts of propagating phase (excitation) or diffusive waves represent signals, or values of logical variables. When fronts interact at the junctions some fronts annihilate or new fronts emerge. They represent a result of the computation. The approach is far from being elegant (compared for instance to the collision-based paradigm~\cite{cbc}) but simple and practical enough to be used by experimental scientists and engineers. This method of constraining the excitable media geometrically has enabled the design of a range of computing devices in the chemical laboratory. This has included logical gates~\cite{toth_showalter_1995,sielewiesiuk_2001}, diodes~\cite{kusimi_1997,dupont_1998,motoike_2003}, counters~\cite{gorecki_2003}, coincidence detectors~\cite{gorecka_2003} and memory~\cite{motoike_2001} all carried out using analogues of the Belousov-Zhabotinsky (BZ) medium.

Until very recently precipitating chemical systems had not been used in unconventional computing. However, previously we demonstrated in chemical laboratory experiments~\cite{adamatzky_delacycostello_XOR} that an {\sc xor} gate can be constructed in the reaction between palladium chloride and Potassium iodide (the palladium processor) via the use of a simple T-junction constructed using a substrate loaded gel. In the reaction precipitation fronts stop propagating when they encounter each other leaving a precipitate free zone between the competiting fronts (competition for the substrate). The ideas was further developed in the construction of logical gates in propagating slime mold~\cite{tsuda_2004} (repulsion of growing patterns).

As discussed in~\cite{kn:ADA05} there are certain (dis)advantages of excitable and precipitating chemical media. Excitable systems are always (assuming unlimited supply of resources and removal of by-products) returning to their original resting state, and thus are capable of processing multiple streams of information. However they do not have static memory. Precipitating systems are disposable (it takes too much energy to reverse precipitation). However precipitating systems always maintain the result of the computation in their long-term memory. So, both types of systems are equally `useful'. 

In the present paper we investigate the possibility of computation in self-constrained reaction-diffusion systems. We develop and analyze a cellular automaton model where the channels are constructed using stationary localizations and computation is implemented by propagating patterns of 
precipitation.

\section{The cellular automaton}

Previously we have proposed a detailed classification of two-dimensional semi-totalistic cellular automaton functions~\cite{kn:AJS05}, which imitates the processes of dissociation and association in spatially extended nonlinear media. Every cell of the automaton array has eight neighbours and takes two states, $0$ and $1$. Cells update their states simultaneously in discrete time, depending on the sum of the states of their neighbours, and by the same cell-state transition rule. A cell in state $0$ takes state $1$ if the number of neighbours in state $1$ belongs to the interval $[\theta_1, \theta_2]$, otherwise the cell remains in the state $0$. A cell in state $1$ continue in state $1$ if the number of its neighbours in state $1$ belongs to interval $[\delta_1, \delta_2]$, otherwise the cell changes the state $0$. The transition parameters satisfy the following condition: $0 \leq \theta_1 \leq \theta_2 \leq 8$ and $0 \leq \delta_1 \leq \delta_2 \leq 8$. This way, the rules can be written as $\theta_1 \theta_2 \delta_1 \delta_2$.

%A cell in state $1$ takes the state $0$ if the number of its neighbours in state $1$ belongs to interval $[\delta_1, \delta_2]$, otherwise the cell takes the state $1$.

\begin{figure}
\centerline{\includegraphics[width=4.2in]{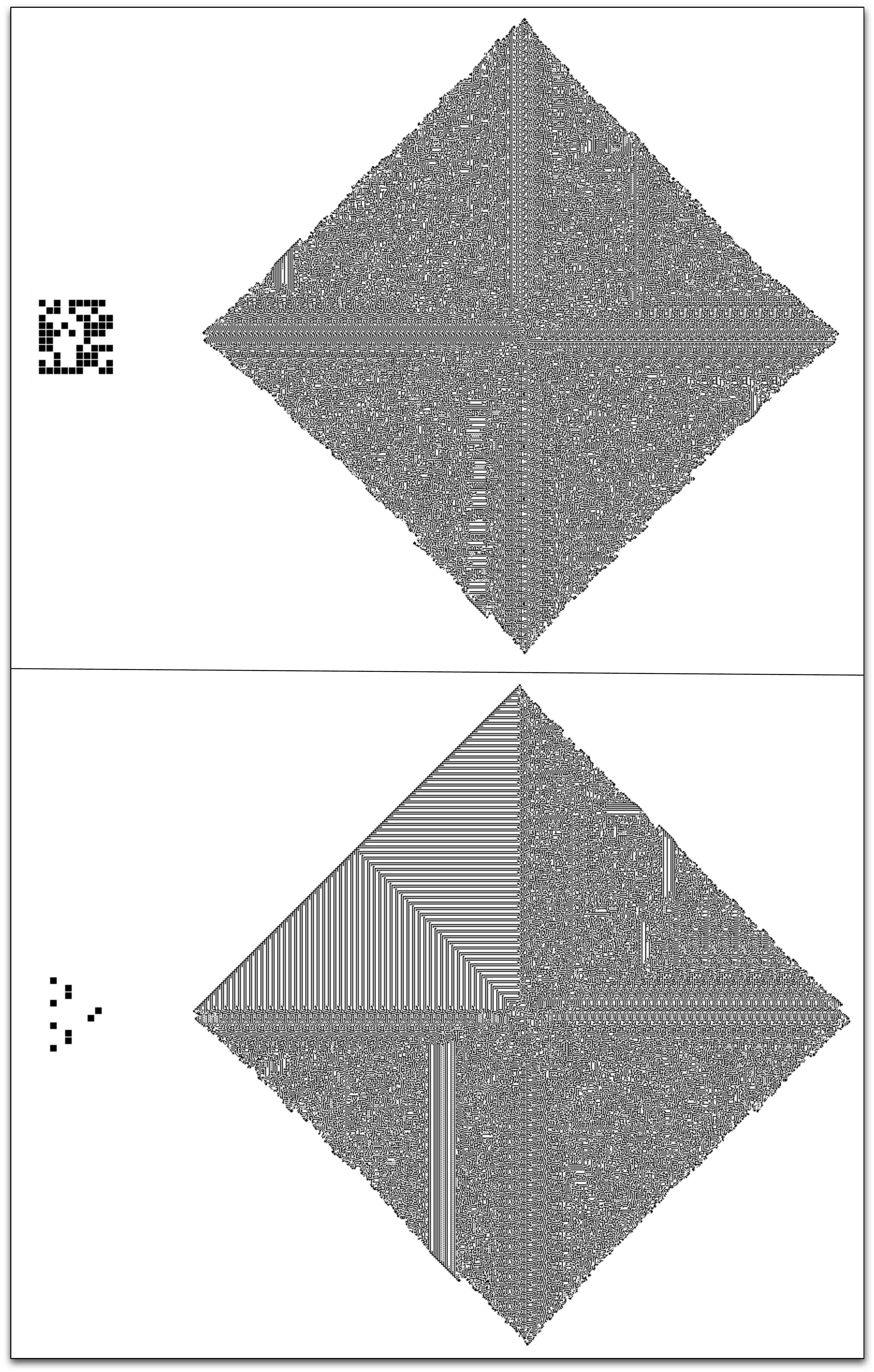}}
\caption{Patterns constructed with the evolution rule 2422. First case show an initial random perturbation into a box of $10 \times 10$ cells, initial density of \%50, evolution in 335 times with 105,456 live cells. Second case show reactions of two gliders with a small oscillator period two, evolution in 360 times with 107,519 live cells.}
\label{2422}
\end{figure}

Such a cell-state transition function can be interpreted as a simple discrete model of a quasi-chemical system with substrate `0' and reagent `1',  where $[\theta_1, \theta_2]$ is an interval of reaction (association or precipitation), and $[\delta_1, \delta_2]$ is an interval of dissociation. These family of rules includes Conway's Game of Life (GoL), when intervals $[\delta_1, \delta_2]$ and $[\theta_1, \theta_2]$ are interpreted as intervals of survival and birth respectively.\footnote{Life Info \url{http://www.pentadecathlon.com/}}

%\begin{figure}
%\centering
%\subfigure[]{\includegraphics[width=0.15\textwidth]{exm1}}
%\subfigure[]{\includegraphics[width=0.8\textwidth]{exm1a}}
%\caption{A random initial perturbation~(a) in $2c22$ medium leads to an unlimitedly growing pattern~(b).}
%\label{example}
%\end{figure}

We have demonstrated that this family of cellular automata is capable of constructing Voronoi diagrams~\cite{kn:ADA05}, information transfer in the form of mobile self-localizations~\cite{kn:JAM} and self-development of sophisticated patterns from minimal seed conditions \cite{kn:AJS05}.

In this way, we have also discovered and verified in computational experiments that a bigger cluster of rules {\it Life} $dc22$, when $d$ and $c$ takes values of 2 to 8 and $d \leq c$ uniquely supports mobile self-localizations (gliders),\footnote{\url{http://uncomp.uwe.ac.uk/genaro/diffusionLife/life_2c22.html}} almost frequently growing in square patterns, and complex patterns combining highly-disordered and ordered zones they are reached by initial perturbations or from reactions of mobile self-localizations~\cite{kn:AJS05}; see some examples in Fig.~\ref{2422}.

An special attention was concentrated to develop computations in the domain of GoL with the also called ``Diffusion Rule'' \cite{kn:AJS05}. However common self-localizations were also involved into the big cluster $dc22$. In this way, a subset of rules called {\it Life} $2c22$ was analyzed independently founding mobile and stationary self-localizations although they were very limited to project possible computations like in GoL. So, a significant problem was stop supernova explosions, but like was reported in \cite{kn:JMM} an special pattern well-known as ``still life'' pattern in GoL, exist in $2c22$. Generally, any CA that grown quickly was very complicated to stop. In this paper, we report as stop these explosions constructing ``walls'' in the rules $2c22$ when $c>3$.

\begin{figure}[th]
\centerline{\includegraphics[width=4.5in]{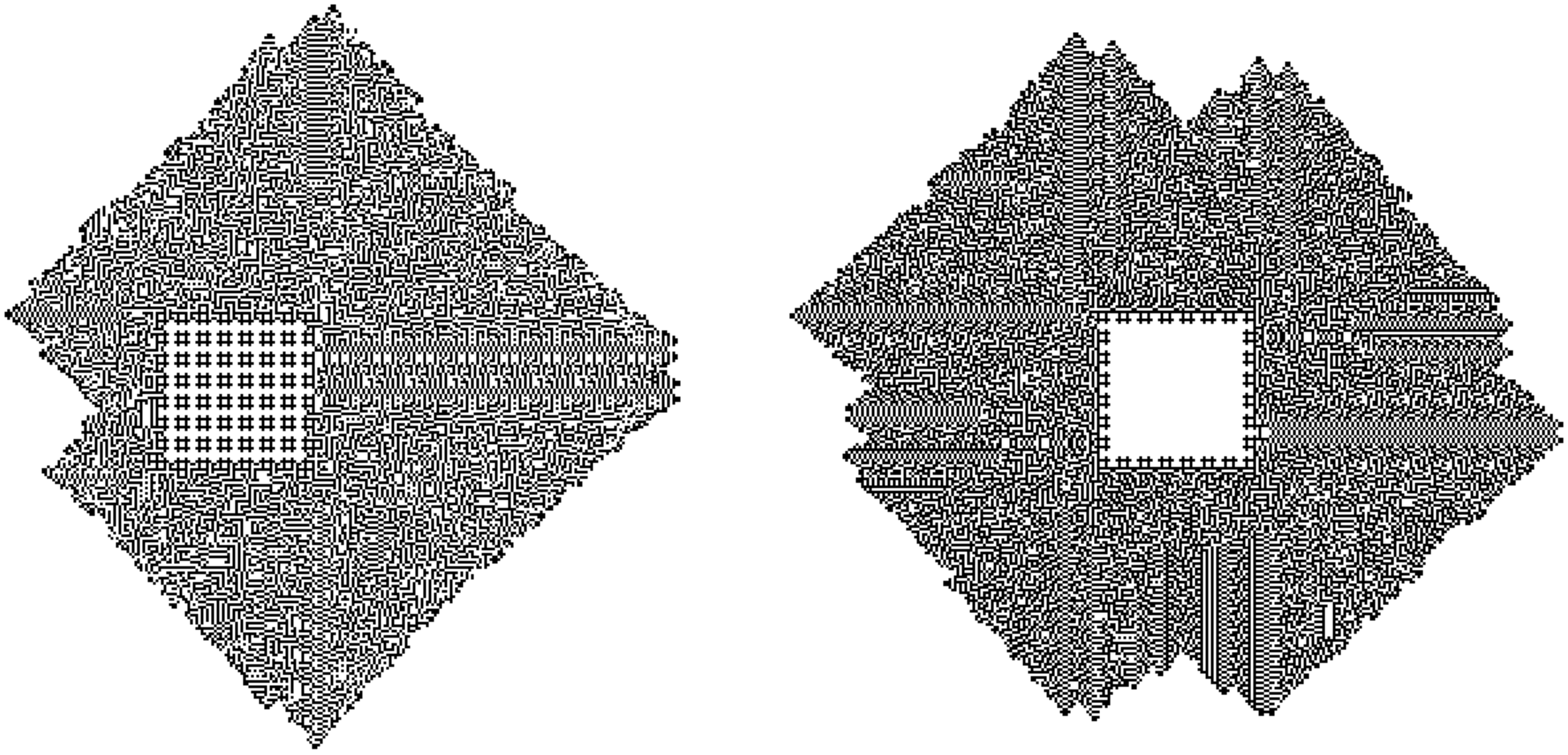}}
\caption{Stationary self-localizations containing unlimited growth, two cases they are displayed. First case show an {\it static universe} that conserve its form even while a chaotic growth is present. Second case show an {\it empty universe} where a chaotic growth is evolving out of this area.}
\label{walls}
\end{figure}

Thus a stationary self-localization is defined into a square of $6 \times 6$ cells. A wall is defined like the volume of this stationary structure plus a cell, i.e., {\it volume} $+ 1$ cell. This construction permit contain any information or perturbation as an {\it static universe} (see Fig.~\ref{walls}). 

Stability with unexcited states in initial conditions with excited states, in general was related to nucleation phenomenon \cite{kn:Grav03}, probabilistic analysis with mean field theory \cite{kn:GV87,kn:Mc90} confirm that these rules $2c22$ when $c>3$ most concentrations of cells in state 1 domain, gradually its curve growth fixing stable fixed points in state 1. But, very small initial densities in state 1 (produced by unstable fixed points) produce a nucleation phenomenon from small perturbations or with reactions of self-localizations. Of course, it is related to minimum noise necessary into the evolution space to produce nucleation like in an excitable chemical reaction \cite{kn:BSG05}.

\begin{figure}[th]
\centering
\includegraphics[width=0.7\textwidth]{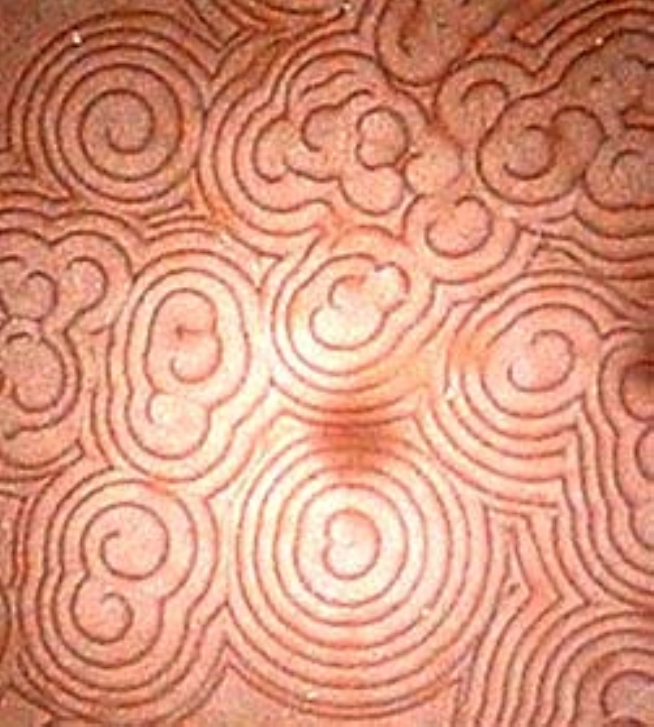}
\caption{Enlarged photo of spiral wave-fronts in precipitating sytems: copper chloride and potassium 
hydroxide. Size is $1 \times 1 mm$~\cite{kn:ADA05}.}
\label{chemexm}
\end{figure}

Complex growing patterns generated by rule $2c22$ do reflect the surprising complexity of some natural precipitating systems, where the apparent spatio-temporal dynamics is morphologically comparable to that of excitable chemical systems. A remarkable example is shown in Fig.~\ref{chemexm}. There we can see complex structures formed when certain concentrations of a simple salt, copper chloride, immobilized in thin gel sheets is reacted with a solution of potassium chloride~\cite{kn:ADA05}.  

In the present paper we construct basic computing primitives using configurations of cellular automata governed by rules $2c22$. We use the propagation of waves in channels of information with stationary and mobile self-localizations. Also, we will see how processing serial and non-serial logic gates. In simulations we have employed {\it Golly} system.\footnote{\url{http://golly.sourceforge.net/}} %and comparative prohabilistic analysis of the rules polynomials using mean field theory~\cite{kn:GV87,kn:Mc90}.

\section{Computing in rules $2c22$}

To implement a universal computation, in terms of Boolean logical gates, in a spatially extended medium one must quantize information in the local propagating patterns. To enable quantization we must prevent any initial configuration from growing unlimitedly. 

In automata rules $2c22$ we have selected the rule $2622$ to made our constructions (clarity in their patterns). Of course, we could use their other four rules but it is not necessary because the other rules were just an adjustment to manipulate the same patterns, potentially all they are candidates to process the same operations and therefore proof its respective logic universality in $2c22$ when $c>3$.

\begin{figure}[th]
\centering
\includegraphics[width=\textwidth]{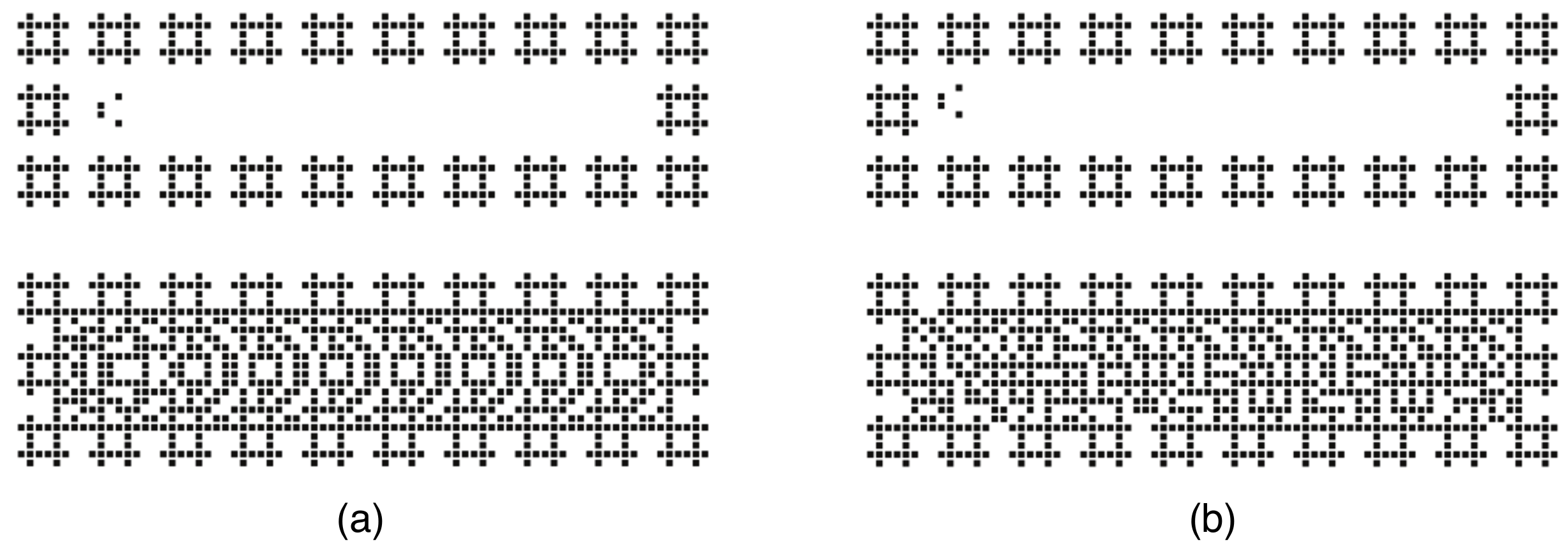}
\caption{Two types of patterns propagation in the information channels (bottom) and their initial configurations stimulated by a mobile self-localization (top).}
\label{wave-comp-1}
\end{figure}

Thus we can make an impenetrable barrier of stationary self-localizations, so-called still life configurations as we saw in previous one section. Primitively, each stationary localization is an intersection of four segments of 1 state, each segment occupies six cells in state 1; the segments intersect so to enclose the $2 \times 2$ domain of cells in state 0. Eventually, a family of patterns there is manipulating its width of channel and positions of mobile self-localizations. But, we have used minimum distance by mobile and stationary self-localization and it was calculated as: {\it interval} = {\it volume still life} $-$ {\it volume glider}. Minimal space between walls limit number of patterns into of them. See exact configurations of the barrier localizations in Fig.~\ref{wave-comp-1}.

Signals are represented by two types of patterns propagating along the channels (Fig.~\ref{wave-comp-1}, bottom). Both patterns are initiated by the same localization (Fig.~\ref{wave-comp-1}, top), consisting of four cells in state 1, but differently positioned. Logical value 0 is represented by regular wave fronts, which are generated by placing localization at the bisector of the channel's longitudinal walls (Fig.~\ref{wave-comp-1}a). To generate less regular patterns (representing value 1) we shift the localization one cell away from the bisector (Fig.~\ref{wave-comp-1}b). 

\begin{figure}[th]
\centering
\includegraphics[width=\textwidth]{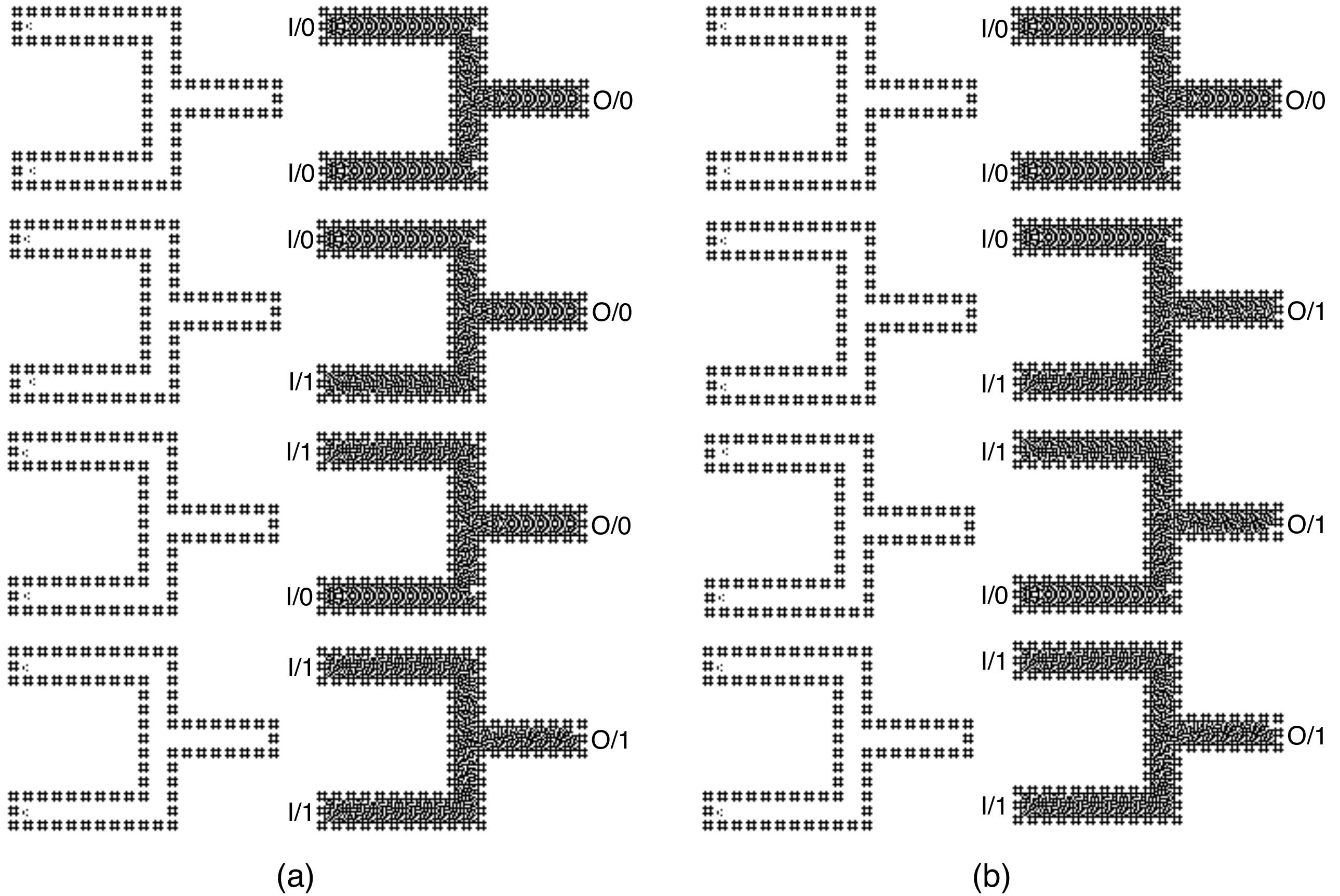}
\caption{Snapshots of cellular automaton configurations, corresponding to implementation of logical gates (a)~{\sc and} and (b)~{\sc or}.}
\label{wave-comp-2}
\end{figure}

Truth values of Boolean variables are represented by these two propagating patterns. Computation occurs when patterns propagating in channels interact at the junctions of the channels. Configurations of cellular automata implementing the logical gates {\sc and} and {\sc or} are shown in Fig.~\ref{wave-comp-2}. Automaton configurations before computation are shown on the left (initial state), besides you can see localizations, representing logical values 0 and 1 depending its position or phase. The configuration at the end of computation, after the patterns have completed their propagation and interaction are shown on the right in both cases.

\begin{figure}
\centerline{\includegraphics[width=4.3in]{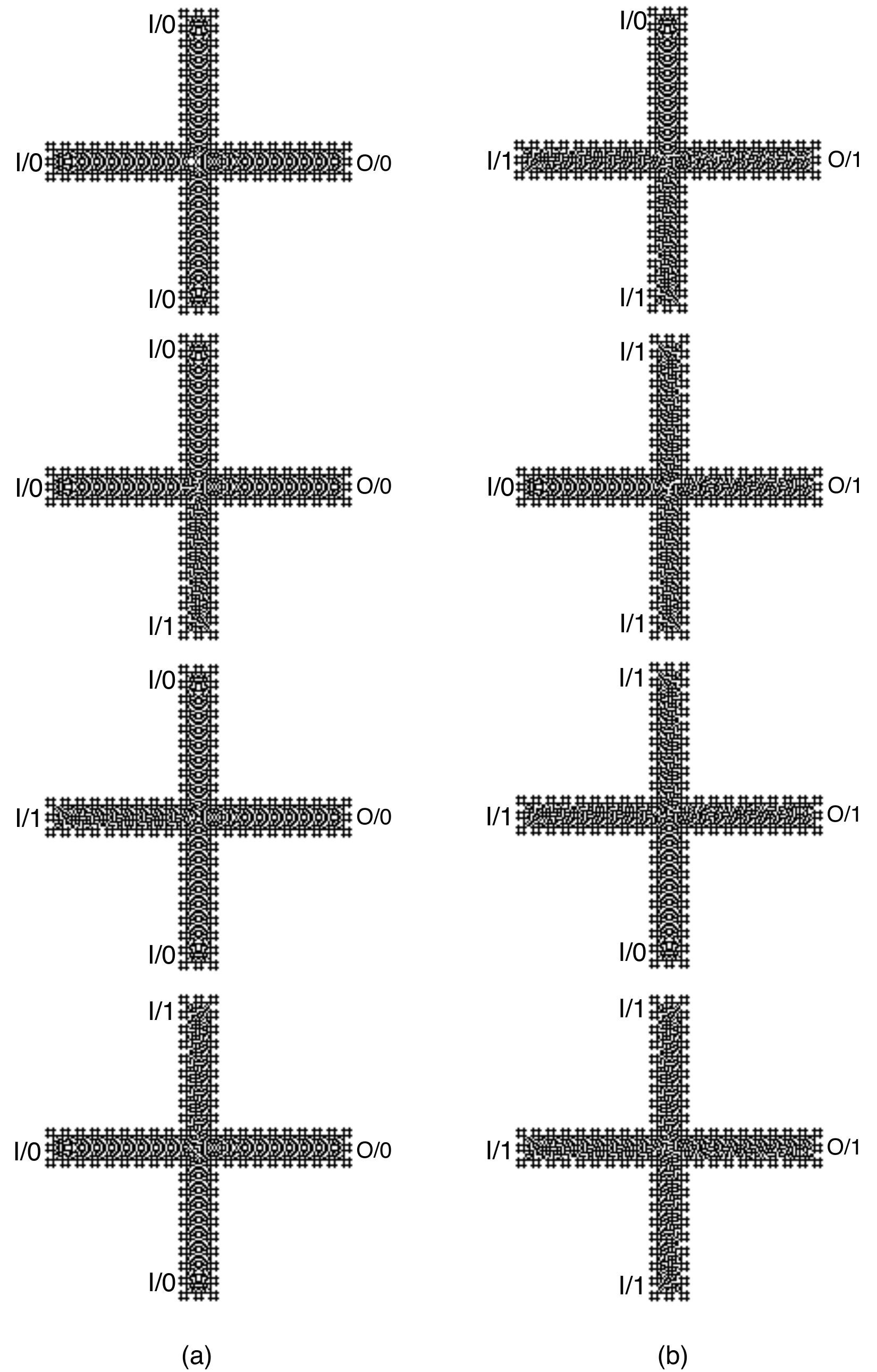}}
\caption{Cellular automaton implementation of {\sc Majority} gate: (a)~all cases corresponding to majority input value I/0, (b)~all cases corresponding to majority input value I/1.}
\label{wave-comp-3}
\end{figure}

%\begin{figure}
%\centering
%\includegraphics[width=\textwidth]{wave-comp-3}
%\caption{Cellular automaton implementation of {\sc Majority} gate:
%(a)~all cases corresponding to output value O/0, 
%(b)~all cases corresponding to output value O/1.}
%\label{wave-comp-3}
%\end{figure}

The channels and junctions, made of stationary localizations, are exactly the same for {\sc and} and {\sc or} gates? The difference is in the exact position of the localization-stimulus representing value 1. Compare configurations corresponding to input $\langle 01 \rangle$ in Fig.~\ref{wave-comp-3}ab. To implement {\sc and} gate the localization representing 1 must be shifted toward the localization representing 0 (shifted North) in Fig.~\ref{wave-comp-3}a. While for the {\sc or} gate the localization representing 1 must shifted away from the localization representing 0 (shifted South).  Such shifts lead to different propagating patterns, representing 1, which either lose ({\sc and} gate) or win ({\sc or} gate) in competition with the propagating pattern, which represents 0. 

Implementation of {\sc Majority} gate is shown in Fig.~\ref{wave-comp-3}. The gate has three inputs: North, West and South channels, and one output: East channel. This gate was constructed when their three propagating patterns hit in the center of the channels thus the final result (O/) is the result of this collisions of waves. Some similarity to construct this gates can be founded with quantum-dot cellular automata \cite{kn:Por99}.

%\clearpage

\section{Discussion}

We presented a cellular automaton model of an abstract quasi-chemical system, where selective and reversible precipitation can occur depending on the local concentration of the species. When the system in a quiescent state is locally disturbed, the disturbance grows unlimitedly. However, if it is surrounded by stationary localizations the disturbance propagates in a controllable manner. We demonstrate that 
such a property of a self-imposed geometrical constraint can be used to implement logical gates, and consequently circuits. The results of the paper give yet more evidence that precipitating chemical systems can be used to construct universal logical circuits. 

Self-restriction of the -- otherwise -- unlimited growth is yet another novel feature of the paper. Previous results, e.g. Banks cellular automata~\cite{banks} or sand pile automata~\cite{goles_1996} have already exploited stationary perturbations, or wires, to guide traveling patterns/signals. However, in their work, signals were rather similar to electrical potential propagating along a wire of conductive material. This way only the rule $2622$ demonstrates computation in the system with unlimited growth self-constrained by stationary localizations. However, the cluster of rules $2c22$ ($c>3$) have the same capacities to support logic universality.

The computation in the $2c22$ medium is based on competition between two types of propagating patterns. We believe the theoretical results discussed will demonstrate their practical value in experimental implementations of geometrically constrained non-linear medium processors, chemical processors and self-growing nano-structures.

\end{document}